\documentclass[prl,twocolumn,groupedaddress,floatfix,amsfonts,amssymb]{revtex4}
\usepackage{amsmath,graphicx}
\usepackage{subfigure}

\newcommand{\la}{\langle}
\newcommand{\ra}{\rangle}
\newcommand{\msd}{\la x^2 \ra}

\begin{document}

\title{``Diffusing diffusivity'': A model for anomalous and ``anomalous yet Brownian'' diffusion}

\author{Mykyta V. Chubynsky}
\email{chubynsky@gmail.com}
\author{Gary W. Slater}
\email{gslater@uottawa.ca}
\affiliation{
Department of Physics, University of Ottawa, 150 Louis-Pasteur, Ottawa, Ontario
K1N 6N5, Canada}

\date{\today}

\begin{abstract}
Wang \textit{et al.} [PNAS 106 (2009) 15160] have found that in several systems the linear time dependence of the mean-square displacement (MSD) of diffusing colloidal particles, typical of normal diffusion, is accompanied by a non-Gaussian displacement distribution (DisD), with roughly exponential tails at short times, a situation they termed ``anomalous yet Brownian'' diffusion. The diversity of systems in which this is observed calls for a generic model. We present such a model where there is ``diffusivity memory'' but no ``direction memory'' in the particle trajectory, and we show that it leads to both a linear MSD and a non-Gaussian DisD at short times. In our model, the diffusivity is undergoing a (perhaps biased) random walk, hence the expression ``diffusing diffusivity''. The DisD is predicted to be exactly exponential at short times if the distribution of diffusivities is itself exponential, but an exponential remains a good fit to the DisD for a variety of diffusivity distributions. Moreover, our generic model can be modified to produce subdiffusion.
\end{abstract}

\pacs{}
\maketitle

In a microscopically homogeneous and rheologically simple (Newtonian) fluid like water, the diffusion of microscopic particles obeys simple laws of Brownian motion known since Einstein~\cite{einstein05}. For instance, the mean-square displacement (MSD) $\msd$ of a particle along a particular direction, $x$, is linear in time $t$,
\begin{equation}
\msd=2Dt,\label{BrownMSD}
\end{equation}
where $D$ is the diffusion constant, or the diffusivity, while the distribution of displacements is Gaussian~\cite{langbook}. In ``crowded'' fluids containing colloidal particles, macromolecules, filaments, etc., the situation can be more complicated and Eq.~(\ref{BrownMSD}) is generally not valid at all times. In many such cases~(see, e.g., Refs.~\cite{danish04, golding06, theriot10, telomeres09, weiss03, franosch13, banks05, weiss04, ernst12, weitz04, weiss09}), experimental data are consistent with
\begin{equation}
\msd\propto t^{\nu},\label{anomMSD}
\end{equation}
where $\nu<1$, over a significant time range. Processes described by Eq.~(\ref{anomMSD}) with $\nu\ne 1$ are called \textit{anomalous diffusion}, more specifically, \textit{subdiffusion} for $\nu<1$.

While much experimental work has concentrated on the MSD, the full displacement distribution (DisD) can be measured using single-particle tracking techniques (SPT)~\cite{danish04, golding06, theriot10, telomeres09, ernst12, weitz04, saxtonSPTrev, metzlerSPT}. In the continuous-time random walk (CTRW) model of anomalous diffusion~\cite{montrollweiss65, bouchaudgeorges90} the DisD is significantly non-Gaussian with a characteristic cusp at $x=0$~\cite{metzlerklafter00}. However, the fractional Brownian motion (fBm) model~\cite{mandelbrot68, jeonmetzler10, milchev11} demonstrates that the combination of anomalous MSD with the normal, Gaussian shape of the DisD is possible. 

On the other hand, it is often tacitly assumed that if the DisD is non-Gaussian, then the factors that cause it to deviate from Gaussian should also make the MSD nonlinear. Recent SPT experiments by Granick's group~\cite{granickPNAS, granickNatMat} show that this is not always the case. Several systems were considered: submicroscopic polystyrene beads on the surface of a lipid bilayer tube~\cite{granickPNAS}, beads in an entangled solution of actin filaments~\cite{granickPNAS}, and liposomes in a nematic solution of aligned actin filaments~\cite{granickNatMat}. In all three systems, the MSD is essentially precisely linear over the whole experimental time range, from $\sim 0.1$ s to a few seconds. Yet, coexisting with this linear MSD is a strongly non-Gaussian DisD, with approximately exponential tails. When a crossover to Gaussian DisD is observed, as in the first system, the linear MSD dependence continues without any peculiarities. This behavior was termed ``anomalous yet Brownian'' diffusion~\cite{granickPNAS}. Very recently, similar behavior was also observed for diffusion of tracer molecules on polymer thin films~\cite{polyfilmindian} and in simulations of a 2D system of discs~\cite{2Ddiskskorean}. Since this is observed in several different cases, it is likely a generic feature of a certain class of systems. The goal of this paper is to show that this may indeed be the case, by proposing a very simple and generic toy model, the \textit{``diffusing diffusivity'' model}, that indeed exhibits this behavior.

First, it is important to realize that while a random walk (RW) with uncorrelated step \textit{directions} has a linear MSD, its DisD will in general be non-Gaussian if the step \textit{lengths} are correlated. Consider for simplicity an unbiased 1D RW with particle displacement $\Delta x_i$ at step $i$ ($i=1,\ldots,N$) and a constant step duration $\Delta t$. The total displacement after $N$ steps is $x_N=\sum_{i=1}^N \Delta x_i$. The MSD is
\begin{equation}
\la x_N^2 \ra = \sum_{i=1}^N \la (\Delta x_i)^2 \ra + 2\sum_{i=1}^{N-1}\sum_{j=i+1}^N \la \Delta x_i \Delta x_j \ra.\label{MSD}
\end{equation}
The second sum in Eq.~(\ref{MSD}) is zero if the steps are uncorrelated ($\la \Delta x_i \Delta x_j \ra=0$); if the ensemble-averaged magnitude of a step is then time-independent (a condition not satisfied for the CTRW), the first sum and thus the whole MSD is proportional to $N$. Note that complete independence of steps is \textit{not necessary} for this to be the case. In particular, if $P(\Delta x_j|\Delta x_i)$ is the conditional probability density for step $j$ to be $\Delta x_j$ given that an earlier step $i$ was $\Delta x_i$, then it is sufficient to have
\begin{equation}
P(-\Delta x_j | \Delta x_i)=P(\Delta x_j | \Delta x_i),\label{condition}
\end{equation}
which is a much weaker condition than complete independence [$P(\Delta x_j | \Delta x_i)=P(\Delta x_j)$]. Equation~(\ref{condition}) means that even if the \textit{direction} of step $i$ is known, the two directions of any future step $j$ are still equiprobable. Thus, the step \textit{directions} are \textit{uncorrelated} (a condition violated by the fBm). However, correlations of the step \textit{lengths} are still allowed.

In fact, such correlations of step magnitudes without correlations of step directions are to be expected in heterogeneous systems where the environment of the diffusing particles changes slowly in space and time. Over length and time scales smaller than those of these heterogeneities, we can describe a local environment approximately by its effective diffusivity. The idea then is to think of a process where on a short time scale particles undergo regular normal diffusion (but with diffusivities different for different particles depending on the local environment), but on a longer time scale, as the environment changes slowly (either on its own, or because the particle moves to a different environment, or both), the diffusivity of each particle changes gradually. Effectively, this leads to long-term correlations between step magnitudes: a long step $\Delta x$ is more likely to be associated with a region with high diffusivity, and then subsequent steps of the same particle are also likely to be longer than average, until the environment (and hence the diffusivity) changes. However, the step directions remain uncorrelated.

In the spirit of the preceding discussion, consider a model in which an ensemble of non-interacting particles diffuse in 1D, each with its own instantaneous diffusion coefficient (or diffusivity) that varies with time. Over a fixed $\Delta t=1$, a specific particle with diffusivity $D_i$ at time step $i$ is displaced by amount $\Delta x_i$ drawn from the Gaussian distribution
\begin{equation}
P(\Delta x_i)=\frac{1}{\sqrt{4\pi D_i}}\exp\left(-\frac{\Delta x_i^2}{4D_i}\right).\label{P1a}
\end{equation}
In the stationary state, the diffusivity distribution is time-independent, and the ensemble-averaged MSD is linear in the time (or the number of steps $N=t/\Delta t$):
\begin{equation}
\la x_N^2 \ra=\sum_{i=1}^N \la \Delta x_i^2 \ra = 2\sum_{i=1}^N \la D_i \ra = 2\la D\ra N.
\end{equation}
On the other hand, the fourth moment of the DisD deviates from its Gaussian value:
\begin{eqnarray}
\la x_N^4\ra-3\la x_N^2\ra^2
&=&12\left(\la D^2\ra-\la D\ra^2\right)N\nonumber\\
& &\hspace{-2cm}+24\sum_{i=1}^{N-1}\sum_{j=i+1}^N \left(\la D_i D_j\ra-\la D_i\ra\la D_j\ra\right).\label{devi4th}
\end{eqnarray}
If the correlator $\la D_i D_j\ra-\la D_i\ra\la D_j\ra$ decays in $\tau_D\gg 1$ steps, then the double sum in Eq.~(\ref{devi4th}) dominates. Assuming $\la D^2\ra-\la D\ra^2\sim\la D\ra^2$, the non-Gaussianity parameter $\alpha_2=\la x^4\ra/3\la x^2\ra^2-1\sim 1$ for $N\lesssim\tau_D$ and $\alpha_2\sim\tau_D/N$ for $N\gg\tau_D$, which only decays as $1/N$. Thus, significant deviations from Gaussianity are expected well above $\tau_D$, especially when looking at the tails of the DisD. This is an important point when interpreting experimental results.

Let us now make specific assumptions about the evolution of $D$ for individual particles. Based on the above, for the DisD to remain non-Gaussian at times $\gg 1$, we need $\tau_D\gg 1$, i.e., $D$ needs to change slowly. Given that this process is expected to be quasirandom in a complex system, we assume that $D$ undergoes a (perhaps biased) RW (``diffusing diffusivity''). Switching now to a continuous time $t$, the equation for evolution of the diffusivity distribution (DifD) $\pi(D;t)$ is then
\begin{eqnarray}
\frac{\partial \pi(D;t)}{\partial t}&=&-\frac{\partial J}{\partial D},\label{difdifbiased}\\
-J&=&\frac{\partial}{\partial D}[d(D) \pi(D;t)]+s(D)\pi(D;t),\label{biasedbc}
\end{eqnarray}
where $d(D)$ is the ``diffusivity of diffusivity'' and $-s(D)$ is the ``force'' biasing the ``diffusion of diffusivity''. Since $D$ cannot be negative or higher than the free-solution diffusivity $D_{\rm max}$, we add reflecting boundary conditions $J=0$ at $D=0$ and $D=D_{\rm max}$. When applied to all $D$, $J(D)=0$ gives the stationary solution of Eq.~(\ref{difdifbiased}), which we denote $\pi(D)$. In what follows, unless stated otherwise, we assume that the system is in this stationary state.

Over times $t\ll\tau_D$ the diffusivity of a particle can be assumed constant. The DisD for an ensemble of particles over such times does not depend on the ``diffusivity diffusion'', but only on $\pi(D)$, and is given by~\cite{granickNatMat, worms09}
\begin{equation}
G(x;t)=\int_0^{D_{\rm max}} \frac{\pi(D)}{2\sqrt{\pi Dt}}\exp\left(-\frac{x^2}{4Dt}\right)\, dD.\label{DDgen}
\end{equation}
Interestingly, for an exponential DifD and with $D_{\rm max}\to\infty$, that is,
\begin{equation}
\pi(D)=\frac{1}{D_0}\exp(-D/D_0),\label{expstat}
\end{equation}
we get~\cite{besselnote}
\begin{equation}
G(x;t)=\frac{1}{2\sqrt{D_0 t}}\exp\left(-\frac{|x|}{\sqrt{D_0 t}}\right).
\end{equation}
In other words, for an exponential distribution of diffusivities the DisD is likewise exactly exponential. Even for finite $D_{\rm max}$, if $D_0\ll D_{\rm max}$, the DisD is still going to be close to exponential if $|x|$ is not too large. 

\begin{figure}
\includegraphics[width=3.3in]{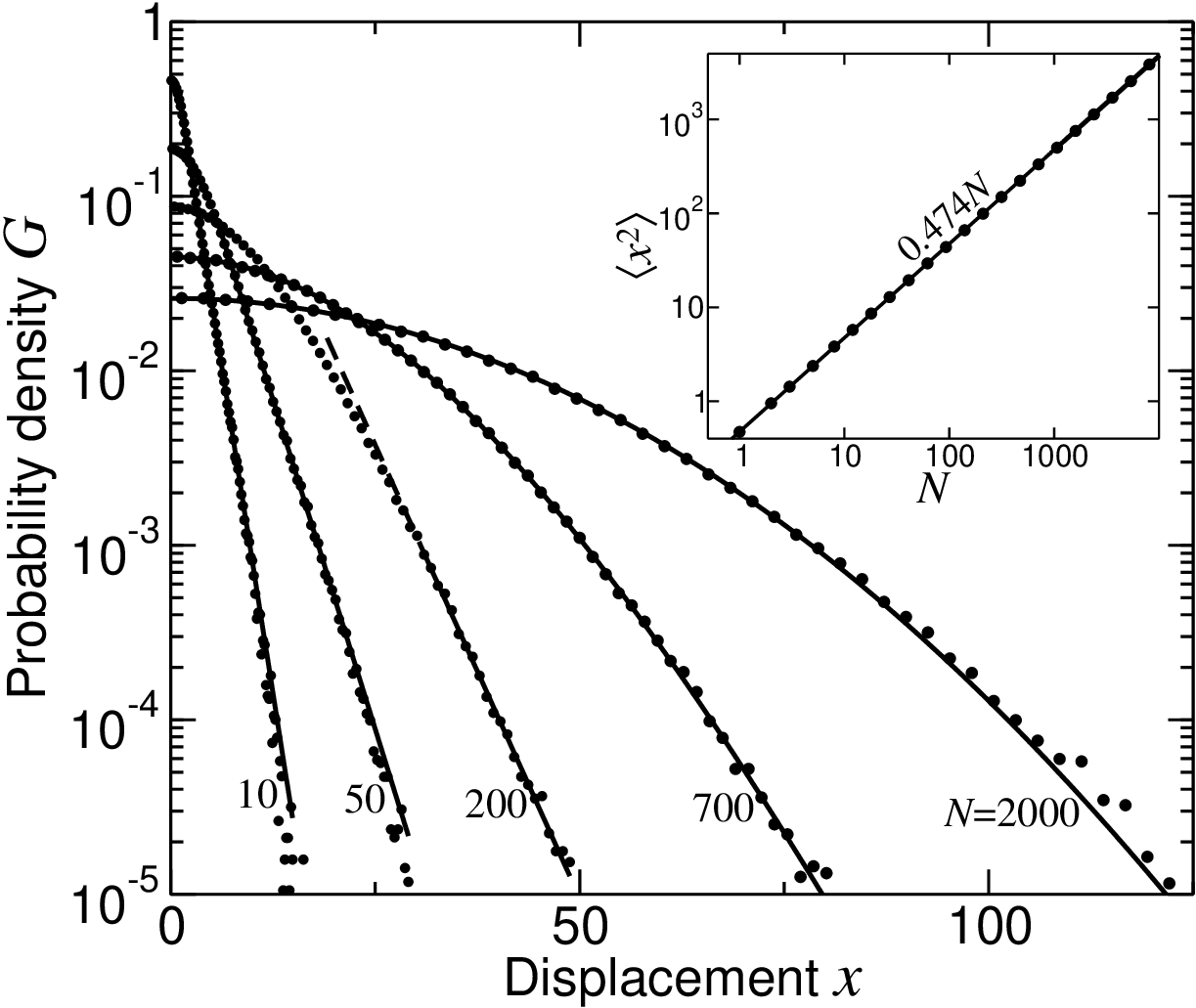}
\caption{The displacement distributions after several different numbers of steps $N$ for the diffusing diffusivity model [Eqs.~(\ref{P1a}), (\ref{difdifbiased}), (\ref{biasedbc})] with $d=0.0025$ and $s=0.01$ simulated as described in the text. The solid line fits are: exponential for the three smallest values of $N$, Gaussian for $N=2000$, and a function $G(x)=A\exp(-B\sqrt{1+(x/X_0)^2})$ interpolating between a Gaussian and an exponential for $N=700$. The inset shows the MSD and a linear fit.\label{fig1}}
\end{figure}

An exponential DifD can be obtained in our model when in Eq.~(\ref{biasedbc}) both $d$ and $s$ are constant, which gives $D_0=d/s$. This is perhaps the simplest reasonable model within our approach (without the bias, the DifD is uniform, which seems unlikely in practice). We simulate this model by a Monte Carlo procedure using the time step $\Delta t=1$, drawing the particle displacement at each step from the Gaussian distribution (\ref{P1a}) and changing the diffusivity $D_i$ at each step drawing the increment from the Gaussian distribution with variance $2d$ shifted by $-s$. Whenever $D$ exits the interval from $0$ to $D_{\rm max}=1$ it is reflected back into this interval. We chose small values $d=0.0025$ and $s=0.01$ so the change of $D$ during one step is likewise small. Drawing the initial diffusivities from the uniform distribution on [0;1], we let them evolve for 1000 steps before taking displacement data, which is sufficient for convergence of the DifD towards that approximately given by Eq.~(\ref{expstat}). Fig.~\ref{fig1} shows that exponential fits to the DisD tails at short $t$ are very successful, despite the fact that $D_0=0.25$ is rather large (the condition $D_0\ll D_{\rm max}=1$ is only weakly satisfied). Note that Wang \textit{et al.}~\cite{granickPNAS} found the observation of ``anomalous yet Brownian'' diffusion on lipid tubules particularly surprising, since the ratio of the effective diffusivity in the system and the free-solution diffusivity was atypically large for systems with anomalous diffusion, about $1/5$. Interestingly, the ratio of the mean diffusivity $\la D\ra\approx D_0$ to $D_{\rm max}$ is even higher in our model.

While the above provides a simple model for ``anomalous yet Brownian'' diffusion, one may certainly argue that the DifD is unlikely to be exponential in all of the diverse cases studied by Wang \textit{et al.}~\cite{granickPNAS, granickNatMat}. We note, however, that the exponential fits of Wang \textit{et al.} are not perfect; therefore, showing that other types of DifD produce DisDs with quasi-exponential tails would go some way towards addressing this concern. For a particular DifD, the integral (\ref{DDgen}) can be done numerically and the asymptotic behavior can also be estimated using the steepest-descent method~\cite{steepdesc}. For instance, for a modified exponential DifD, $\pi(D)\sim D^{\alpha}\exp(-D/D_0)$, there is a power-law prefactor in the form of the tail of the DisD, $G(x;t)\sim t^{-(\alpha+1)/2}|x|^\alpha\exp(-|x|/\sqrt{D_0 t})$. This prefactor is just a logarithmic correction on a semi-logarithmic plot, and if $|\alpha|$ is small (say, below 2), fitting with an exponential over several decades in $G$ is still adequate. It is also interesting to note that when the DifD is a stretched or compressed exponential, $\pi(D)\sim\exp[-(D/D_0)^\beta]$, $G(x;t)$ is likewise a stretched or compressed exponential (up to a power-law prefactor) but with the exponent $\gamma=2\beta/(\beta+1)$, which is always closer to unity than $\beta$. For instance, for a Gaussian DifD ($\beta=2$), $\gamma=4/3$, and thus the tails are closer to being exponential than Gaussian. In fact, it is only for $\beta\to\infty$, when the DifD approaches a step function, that $\gamma\to 2$ and the tails become Gaussian-like.

\begin{figure}
\includegraphics[width=3.3in]{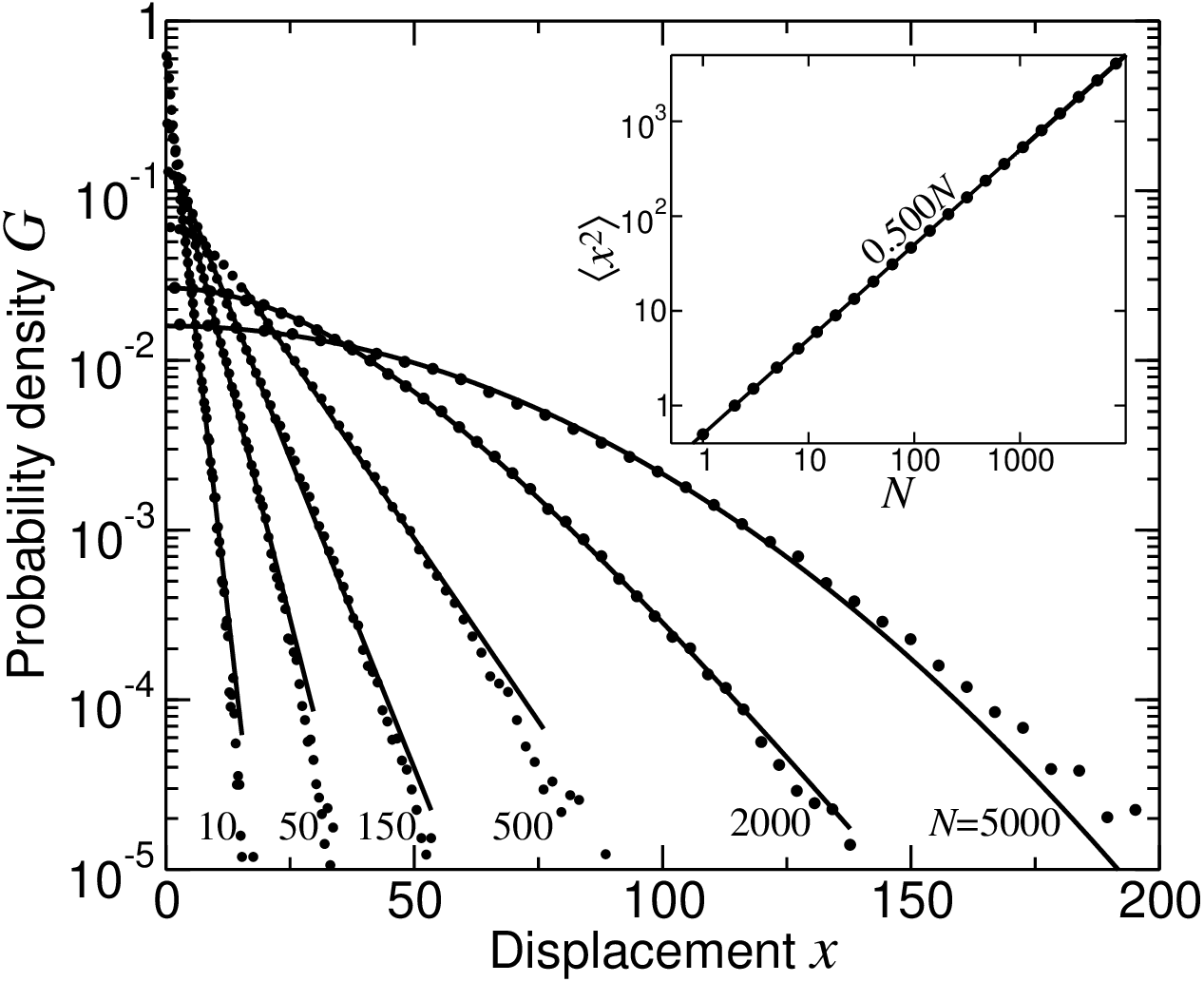}
\caption{The displacement distributions after several different numbers of steps $N$ for the diffusing diffusivity model with coupling to particle displacement with $d_0=10^{-5}$ and $f=10^{-3}$ in Eq.~(\ref{deltacoupled}), simulated as described in the text. The solid line fits are: exponential for the four smallest values of $N$, Gaussian for $N=5000$, and a function $G(x)=A\exp(-B\sqrt{1+(x/X_0)^2})$ interpolating between a Gaussian and an exponential for $N=2000$. The inset shows the MSD and a linear fit.\label{fig2}}
\end{figure}

Another case of near-exponential tails for an even more drastically non-exponential DifD is provided by a physically motivated variant of the diffusing diffusivity model with diffusivity changes coupled to particle displacement. Instead of introducing the bias $s$ artificially, let us make the simplest assumption that the changes in the particle environment alone would lead to a constant ``diffusivity of diffusivity'' without bias. As for the motion of the particle itself between different environments, it is logical to assume that the longer the size of the particle step, the larger the typical diffusivity change during that step. With this in mind, consider the algorithm in which for each step $\Delta x_i$ the associated diffusivity change $\Delta D_i=D_{i+1}-D_i$ is Gaussian with half-variance
\begin{equation}
d=d_0+f(\Delta x_i)^2,\label{deltacoupled}
\end{equation}
where $d_0$ and $f$ are constants. The first term reflects the random fluctuations of the local environment and the second term is due to the particle's moving between different environments. As the mean-square step size is itself proportional to the diffusivity ($\la\Delta x_i^2\ra=2D_i$), this is approximately equivalent to having $d(D)=d_0+2fD$ and $s(D)=0$ in Eq.~(\ref{difdifbiased}). The stationary DifD is then (still using $D_{\rm max}=1$)
\begin{equation}
\pi(D)=\frac{2f}{\ln (1+2f/d_0)}\times \frac{1}{d_0+2fD},\label{coupledstat}
\end{equation}
quite different from Eq.~(\ref{expstat}). Yet, the resulting DisD $G(x,t)$ can still be fitted with exponentials at short times over a significant region, as shown in Fig.~\ref{fig2}. For this figure we have used $d_0=5\times 10^{-5}$ and $f=10^{-3}$, for which Eq.~(\ref{coupledstat}) gives $\la D\ra\approx 0.244$, about the same as in the first version of the model. As before, we let the diffusivities evolve, this time for 2000 steps, before collecting data. The success of the exponential fits is due to the crossover from the approximately power-law DisD at small displacements to the approximately Gaussian DisD above $\sim\sqrt{D_{\rm max}t}$; on the semilog scale this corresponds to regions of different convexity with an inflection point between them. On the other hand, it is clear that for $\la D\ra$ even closer to $D_{\rm max}$ exponential fits will not be as good, since the crossover to the Gaussian distribution at $\sim\sqrt{D_{\rm max}t}$ will be too close to the root-mean-square displacement $\sim\sqrt{\la D\ra t}$. We have checked, in particular, that this is indeed the case when the stationary DifD is uniform on $[0;D_{\rm max}]$, with $\la D\ra=D_{\rm max}/2$. Therefore, the non-exponential DisDs seen in Ref.~\cite{nonexpDisD}, where the observed diffusivity is  $>70$\% of $D_{\rm max}$, are entirely expected. There may be other situations where exponential fits fail, for instance, distinctly bimodal DifDs~\cite{supercooled13}.

\begin{figure}
\includegraphics[width=3.3in]{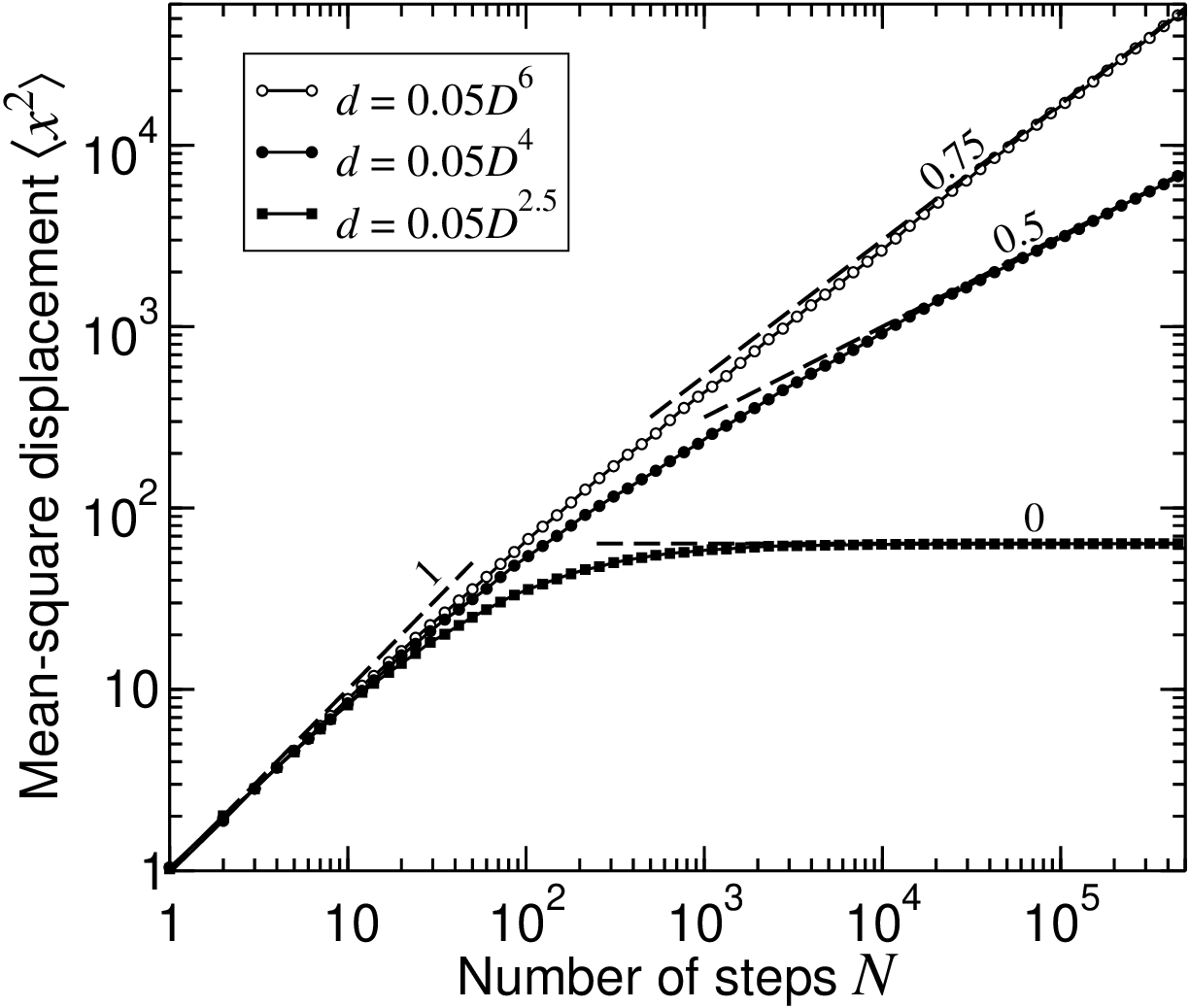}
\caption{The MSD as a function of the number of steps $N$ for the diffusing diffusivity model [Eqs.~(\ref{P1a}), (\ref{difdifbiased}), (\ref{biasedbc})] with $s=0$ and $d$ as specified in the legend. The dashed lines are power laws with the exponents indicated.\label{fig3}}
\end{figure}

It is also interesting that the same general approach can be used to produce \textit{subdiffusion}. Let $s(D)=0$ and $d(D)\propto D^b$ in Eq.~(\ref{difdifbiased}), where $b>3$ is a constant. In this case, there is no stationary solution, except for the trivial $\pi(D)=\delta(D)$. Instead, there is a quasi-stationary solution of the form $\pi(D;t)=t^c f(Dt^c)$, with $c=1/(b-2)$. This solution corresponds to ageing with $\la D\ra$ decreasing gradually to zero as $t^{-c}$ and the corresponding anomalous diffusion exponent therefore being $\nu=1-c=(b-3)/(b-2).$ Results of simulations of this model for several values of $b$ are shown in Fig.~\ref{fig3}. In all cases, the initial DifD is chosen to be uniform on the interval [0;1], and the initial normal diffusion stage therefore corresponds to $\la D\ra=1/2$. For $b>3$, once the quasi-stationary distribution is established, anomalous diffusion sets in. For $2<b<3$, $\la D\ra$ decays faster than $t^{-1}$, and the MSD instead approaches a constant --- the particle remains ``trapped' forever.

To summarize, we have considered a simple 1D toy model of ``anomalous yet Brownian'' diffusion. Recognizing the fact that the linear MSD combined with a non-Gaussian DisD should be observed for random walks without direction memory, but with ``diffusivity memory'', we have assumed that the instantaneous effective diffusion coefficient of a particle changes gradually at random, or ``diffuses'' (with or without bias) --- hence our expression ``diffusing diffusivity''. The short-time DisD is determined solely by the stationary DifD, which can be varied within the framework of the model. This DisD is exactly exponential when the DifD is exponential, but an exponential remains a good fit to a significant part of the tail of the DisD for a variety of DifDs, which may explain the experimental results. The same approach can produce subdiffusion, which may potentially provide yet another possible route to subdiffusion in addition to the CTRW and the fBm, although the peculiar power-law dependence of the ``diffusivity of the diffusivity'' required needs to be justified on physical grounds.

It should be mentioned that, just like the CTRW, our approach is ``mean-field'' since random diffusivity changes neglect the possibility of returning to the same environment. This is a better approximation in higher dimensions and for environments changing sufficiently rapidly on their own. To what extent the conclusions are modified, in particular, in the least favorable case of a static diffusivity distribution in 1D will be a subject of future studies. Also, while we have assumed a gradual change of the diffusivity, we expect the results to be applicable qualitatively to the case when there are sharp boundaries between regions of different diffusivities, as long as the regions themselves are large enough, so they take a long time to traverse. Again, this will be tested in the future.

Finally, we note the similarity of our model to some models of market price fluctuations (see, e.g.,~\cite{stanley00}), although the details differ.

We would like to thank S.~Granick (UIUC), M.~Rubinstein (UNC), Ming Guo (Harvard), A.~Cherstvy (Potsdam), as well as the current and former members of our research group, for useful discussions. This research was supported by the Natural Sciences and Engineering Research Council of Canada (NSERC) and the University of Ottawa.

\end{document}